\documentclass{llncs}
\usepackage{graphicx}
\usepackage{setspace}
\usepackage{csquotes}
\usepackage{booktabs}
\usepackage{tabularx}
\usepackage{listings}
\usepackage{comment}
\usepackage{enumerate}
\usepackage{float}
\usepackage{xcolor}
\usepackage{nameref}
\usepackage{hyperref}
\usepackage{multirow}
\usepackage{makecell}
\usepackage{diagbox}
\usepackage{longtable}
\usepackage{color,soul}

\begin{document}
\title{Artifact Validity in Design Science Research (DSR): A Comparative Analysis of Three Influential Frameworks\thanks{This is an author's version. The final authenticated version is available in the 20th International Conference on Design Science Research in Information Systems and Technology (DESRIST 2025) proceedings.}}

% This is an extended version of the paper available in the 20th International Conference on Design Science Research in Information Systems and Technology (DESRIST 2025) proceedings. All extensions are available in the appendix. All extensions are \ext{marked} \mext{pink}.

%The final authenticated version is available online at https://doi.org/.....

%how to color text: https://tex.stackexchange.com/questions/260566/how-to-reliably-switch-the-highlighting-color-with-soul   and: https://www.overleaf.com/learn/latex/Using_colors_in_LaTeX

\author{Sylvana Kroop}
\institute{University of Vienna, Faculty of Philosophy and Education\\
\email{sylvana.kroop@univie.ac.at}
}%

\maketitle

\begin{abstract}
Although the methodology of Design Science Research (DSR) is playing an increasingly important role with the emergence of the ‘sciences of the artificial’, the validity of the resulting artifacts is occasionally questioned. This paper compares three influential DSR frameworks to assess their support for artifact validity. Using five essential validity types (instrument validity, technical validity, design validity, purpose validity and generalization), the qualitative analysis reveals that while purpose validity is explicitly emphasized, instrument and design validity remain the least developed. Their implicit treatment in all frameworks poses a risk of overlooked validation, and the absence of mandatory instrument validity can lead to invalid artifacts, threatening research credibility. Beyond these findings, the paper contributes (a) a comparative overview of each framework’s strengths and weaknesses and (b) a revised DSR framework incorporating all five validity types with definitions and examples. This ensures systematic artifact evaluation and improvement, reinforcing the rigor of DSR. 
\end{abstract}
\section{Introduction}
\label{sec:intro}
With the rise of the ‘sciences of the artificial’~\cite{simon_sciences_1996}, Design Science Research (DSR) methodology plays an increasingly important role and is gaining ground in higher education institutions much faster than in previous years and decades. With Hevner~\cite{hevner_design_2004,hevner_design_2010}, DSR has become widespread and popular in the field of information technology, but also through intensive efforts in German-speaking countries, especially in business informatics~\cite{osterle_memorandum_2010}, as it focuses on the creation of novel artifacts. As technically oriented disciplines often permeate all areas of life and play an important supporting role in all subject disciplines, the DSR methodology can be applied to the development of new artifacts of all kinds, from software to physical tools such as printed teaching materials or human body prostheses. 

Although it seems to have been clear among DSR experts for decades what DSR can and cannot achieve, e.g.~\cite{nunamaker_systems_1990,march_design_1995,simon_sciences_1996,gregor_building_2009}, this discussion has only just begun in the context of the education, not only for doctoral students but also for master's students, e.g.~\cite{benner-wickner_leitfaden_2020,knauss_constructive_2021,winter_teaching_2021,hevner_proficiency_2023,schlimbach_teaching_2023}. Frequently asked questions are, for example: How complex can or must a DSR-based research design be in order to be mastered within the framework of a master's thesis? Is it necessary to validate the generalizability of the developed artifact? Is it sufficient that the effectiveness of the artifact is evaluated solely in the context of my company? As DSR becomes more widely used, doubts arise about the validity of the resulting artifacts. 

De Sordi et al. (2020) conducted a content analysis of 152 articles to examine the longitudinal development of DSR projects and the types of artifacts involved. The results suggest that the use of DSR has grown rapidly over the years and that this growth is likely to continue in the future. However, the text further states that 86 \% of DSR artifact evaluations are unrealistic, which means that the criteria used to evaluate the artifact are not practical or feasible in real-world scenarios. The authors call for research to make Design Science Research (DSR) more accessible and less confusing as authors, reviewers, and editors struggle to understand and follow DSR guidelines~\cite{de_sordi_design_2020}. This may indeed be difficult, especially for reviewers who grew up in a world where 'truth' was only found in the nature of descriptive research. 

This paper addresses the examination and comparison of three commonly used DSR frameworks, namely Hevner et al.~\cite{hevner_design_2004}, Peffers et al.~\cite{peffers_design_2007}, and an integrated framework combining Österle et al.~\cite{osterle_memorandum_2010} and Benner-Wickner et al.~\cite{benner-wickner_leitfaden_2020}, and the question: to what extent do these three influential design science research (DSR) frameworks support the validity of artifacts? The fundamental question arises as to which types of validity are essential for the evaluation of an artifact. And the resulting question of how a DSR framework can be improved to emphasize the need for a thorough evaluation of an artifact's validity.

\section{Methodology}
\label{sec:method}

In order to compare the three DSR frameworks with the question of the extent to which they support the validity of artifacts, five validity types were established a priori, which were derived both from experience in the assessment and supervision of DSR-based master's theses and from the basic literature~\cite{cook_quasi-experimentation_1979,shadish_experimental_2002,messick_validity_1989,messick_validity_1994,cronbach_construct_1955}. 

\begin{figure}[htb!]
    \centering
    \includegraphics[width=1.0\textwidth]{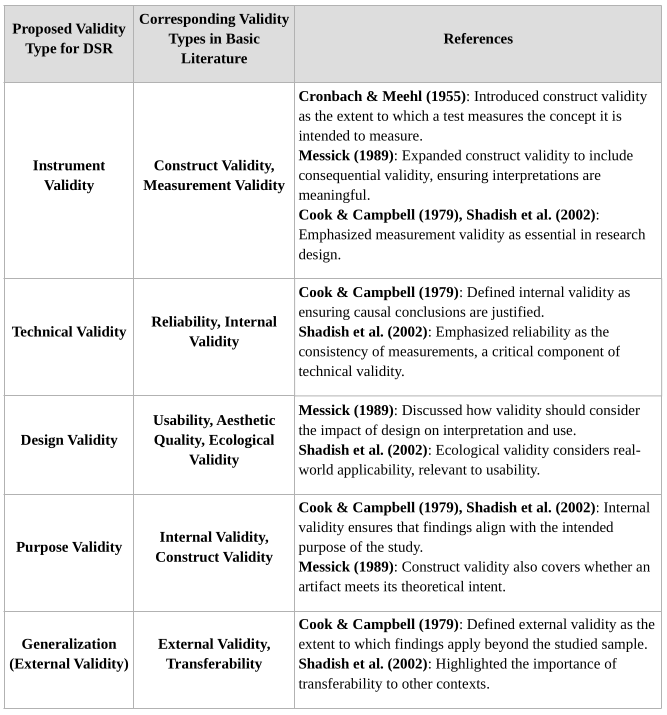}
    \caption{Type of validity categories used a priori for qualitative content analysis}
    \label{fig:tabcategories}
\end{figure}

Each of the five validity types (instruments validity, technical validity, design validity, purpose validity and generalization) proposed for the artifact evaluation in DSR corresponds to several well-known scientific validity concepts, as shown in the overview table in Figure~\ref{fig:tabcategories}. 

The five validity types were then defined and demonstrated using examples for application in the context of design science research (DSR), see Sect.~\ref{sec:valtyp}. Scoring Guidelines for the Five Validity Types were designed, see Sect.~\ref{sec:scoreguide}, and applied in the comparative evaluation of the three DSR frameworks, see Chap.~\ref{sec:discussion}. The resulting scores of the the comparative evaluation were than converted into a heat map, which serves as a final, summary overview, see Chap.~\ref{sec:summarized}.

The comparative evaluation is essentially based on a qualitative content analysis according to the guidelines of Kuckartz \& Rädiker~\cite{kuckartz_qualitative_2022}. The core literature on the three DSR frameworks examined was primarily the subject of the investigation in order to answer the central research question: To what extent do these three influential frameworks of Design Science Research (DSR) support the validity of artifacts?

The three DSR frameworks were selected due to their popularity among master's students. Based on my ongoing supervision of DSR-based theses - especially at the Ferdinand Porsche FernFH in Austria - I’ve observed frequent use of these frameworks in business informatics and IT. However, the revised framework in this paper also enables broader comparisons with other DSR approaches.

\section{Related Work}
\label{sec:rel}

\begin{figure}[htb!]
    \centering
    \includegraphics[width=1.0\textwidth]{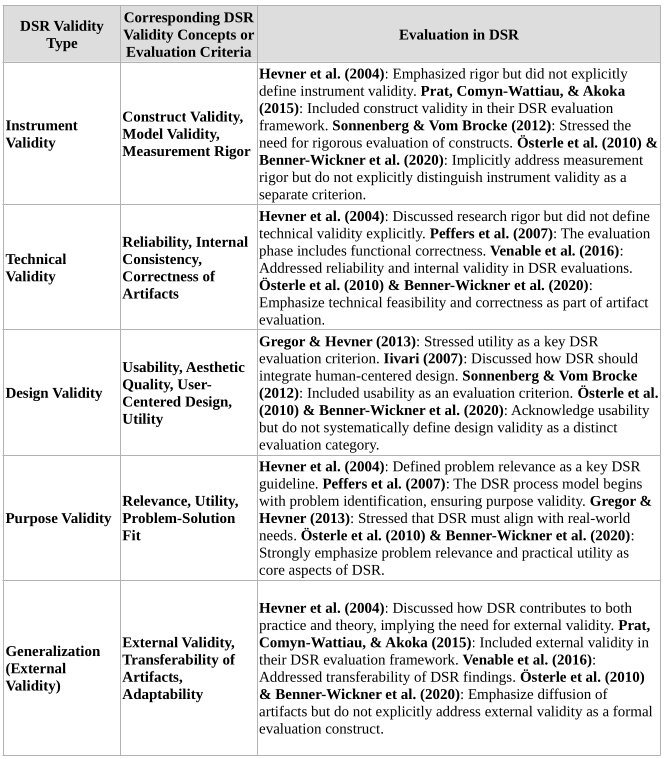}
    \caption{Type of validity categories, aligned with frequently cited DSR literature on artifact evaluation}
    \label{fig:tabrel}
\end{figure}

Similar work comparing the three selected DSR frameworks in terms of support for artifact validity has not yet been conducted. The overview table in Figure~\ref{fig:tabrel} shows how frequently cited literature on artifact evaluation, and other extensive literature, e.g.~\cite{hevner_design_2004,hevner_design_2010,hevner_transparency_2024,chatterjee_typology_2024,hutchison_evaluations_2012,gregor_building_2009,venable_feds_2016,vaishnavi_design_2004,iivari_paradigmatic_2007,iivari_twelve_2010,prat_taxonomy_2015,kuechler_theory_2008,vijay_k_vaishnavi_design_2015,vom_brocke_introduction_2020,brocke_accumulation_2020,benner-wickner_leitfaden_2020,ralph_empirical_2021,osterle_memorandum_2010,peffers_design_2007}, relate to the five validity types applied here. 

In particular, 'instrument validity' as a mandatory prerequisite for a valid artifact (see Sect.~\ref{sec:flextyp}) is difficult to find in all three DSR frameworks examined here. Sonnenberg \& vom Brocke~\cite{hutchison_evaluations_2012} is one of the few studies that clearly points out the need to evaluate the design and the construct separately and in interaction with the problem to be evaluated. This can be equated with the need for 'instrument validity'. Although this model increases the basic evaluation effort, it ensures potentially valid artifacts in the end. Nevertheless, there has been a lack of consideration and differentiation of the five types of validity clearly defined here as an integral part of a comprehensive DSR framework. 

\section{Revised DSR Framework focused on Artifact Validity}
\label{sec:eval}

\subsection{Revised DSR Framework}
\label{sec:revframe}

\begin{figure}[htb!]
    \centering
    \includegraphics[width=1.0\textwidth]{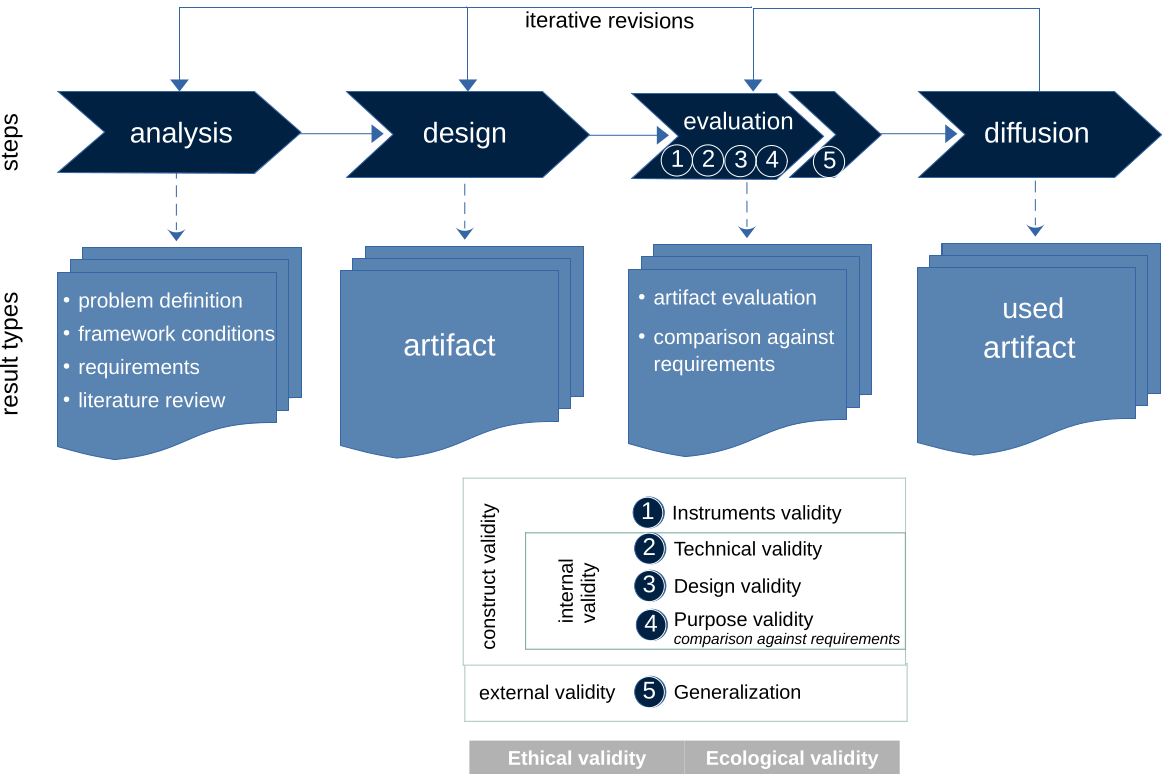}
    \caption{Revised DSR framework focused on artifact validity, based on the integrated DSR framework combining Österle et al.~\cite{osterle_memorandum_2010} and Benner-Wickner et al.~\cite{benner-wickner_leitfaden_2020} }
    \label{fig:valid}
\end{figure}

Compared to Hevner et.al~\cite{hevner_design_2004} and Peffers et.al.~\cite{peffers_design_2007}, the DSR framework, originally developed and popularized by Österle et al.~\cite{osterle_memorandum_2010} and later put into a graphical form by Benner-Wickner et al.~\cite{benner-wickner_leitfaden_2020}, is perhaps less well known because it is only available in German. However, due to its simplicity, it is rather popular among master students in German-speaking countries.
Figure~\ref{fig:valid} is based on this - I call it - "integrated DSR framework", combining Österle et al.~\cite{osterle_memorandum_2010} and Benner-Wickner et al.~\cite{benner-wickner_leitfaden_2020}. 

The revision shown in Figure~\ref{fig:valid} is the extension made in the evaluation step to include the five proposed validity types. The five types of validity were harmonized with the main types of validity (construct, internal and external validity), which have been known and used for decades, see Figure~\ref{fig:tabcategories}. Furthermore, the ethical and ecological validity were pointed out, which should be addressed as standalone measurement instruments, but which would require a separate study. The five validity types are defined in more detail in Sect.~\ref{sec:valtyp} and demonstrated using examples. 

\subsection{Validity Types}   
\label{sec:valtyp}
Although the rigor required in design science research (DSR) implicitly encompasses all five types of validity proposed here, none of the three compared frameworks explicitly refer to or differentiate between these fundamentally necessary types of validity. Therefore, they can easily be overlooked by students who rely exclusively on one of these popular DSR frameworks in their master's thesis. To ensure artifact validity, each validity type is defined in its core below, accompanied by a key question and examples, and discussed further in chap.~\ref{sec:discussion}.

\textbf{Instruments validity} ensures the evaluation instruments and metrics used to assess the artifact are reliable, accurate, and aligned with the constructs they are intended to measure. The KEY QUESTION is: Are the evaluation tools (e.g., surveys, log files, usability tests, experimental setups) valid and capable of accurately measuring what they are intended to measure?

EXAMPLE from everyday situations: When designing or using a thermometer for body temperature, it should accurately measure body temperature and not something else (e.g., room temperature or humidity). If a thermometer mistakenly reacts to the air temperature rather than body heat, it \textit{lacks instruments validity} because it is not measuring what it is supposed to.

EXAMPLE of a DSR application: If a learning analytics dashboard is designed to track student engagement, \textit{instruments validity} ensures that the selected metrics (e.g., time-on-task, login frequency) actually measure engagement rather than unrelated factors (e.g., technical errors causing inactivity).

\textbf{Technical validity} ensures the artifact performs as intended without bugs or glitches, focusing on technical reliability and functional correctness. The KEY QUESTION is: Does the artifact’s technical functionality perform as intended and without problems?

EXAMPLE from everyday situations: A car’s brakes must work correctly every time you press the pedal. If they function sometimes but fail at other times, the braking system \textit{lacks technical validity} because it is unreliable and does not consistently perform its intended function.

EXAMPLE of a DSR application: A machine learning-based fraud detection system must be completely free of critical bugs, biases, or instability issues. If the system produces false fraud alerts due to faulty code, misclassifies legitimate transactions because of biased training data, or crashes under high transaction loads, it \textit{lacks technical validity}. Such errors can lead to financial losses, customer distrust, and regulatory penalties, making error-free performance essential for reliable fraud detection.

\textbf{Design validity} evaluates the artifact’s design from a subjective and aesthetic perspective, as well as its alignment with user expectations and contextual relevance. The KEY QUESTION is: Does the artifact exhibit good style, taste, and elegance, making it aesthetically pleasing and intuitive to users?

EXAMPLE from everyday situations: A chair might technically function (it allows sitting), but if it is uncomfortable, ugly, or too complex to use, it \textit{lacks design validity}. A well-designed chair should be comfortable, aesthetically pleasing, and easy to sit on.

EXAMPLE of a DSR application: A finance dashboard for managers must not only provide accurate financial data but also be designed with clarity, usability, and efficiency in mind. If the dashboard is cluttered, difficult to navigate, or visually overwhelming, it \textit{lacks design validity}, as poor design can hinder decision-making even when the data is correct. A well-designed dashboard enhances comprehension, streamlines analysis, and enables managers to make informed decisions quickly.

\textbf{Purpose validity} determines whether the artifact achieves its intended purpose by effectively solving the targeted problem with the defined requirements. The KEY QUESTION is: Does the artifact fulfill the intended goals (does it meet the defined requirements), and are the observed results attributable to the artifact itself (and not to confounding factors)?

EXAMPLE from everyday situations: A parachute’s purpose is to slow down a person’s fall. If a parachute doesn’t open, opens too late, or fails to slow the fall, it \textit{lacks purpose validity}, even if it was well-designed and made of high-quality materials.

EXAMPLE of a DSR application: A waste tracking app designed to help customers reduce landfill waste must be validated by measuring actual waste reduction against initial goals (e.g., a significant decrease in non-recyclable waste disposal per user). Even if the app correctly logs waste disposal data (\textit{technical validity}), features a well-structured and visually appealing interface (\textit{design validity}), and uses reliable tracking metrics (\textit{instruments validity}), it \textit{lacks purpose validity} if users do not actually reduce their waste. This failure could result from ineffective behavioral nudges, lack of actionable insights, or poor integration with waste disposal services, meaning the app does not fulfill its intended goal despite functioning as designed.

\textbf{Generalization (external validity)} assesses whether the performance and effectiveness of the artifact are transferable to other contexts, populations or environments. The KEY QUESTION is: Can the success of the artifact be repeated in other environments with similar results? 

EXAMPLE from everyday situations: A good (universal) cell phone charger should work across multiple cell phone brands and models, not just for one specific device. If a charger only works for one phone and fails for others, it lacks generalization validity.

EXAMPLE of a DSR application: A predictive maintenance model developed for automobile engines may not perform well in other contexts, such as with aircraft engines or manufacturing machinery. Without testing the model in these different domains, it lacks generalization (external validity), meaning its effectiveness beyond the original domain remains uncertain.
\\

The validity framework represents a reconceptualization and a new contribution to DSR methodology. Although the five types of validity proposed and defined here have been aligned with more traditional notions of validity (see Figure~\ref{fig:tabcategories}), they are not simply an adaptation or extension of their use in the natural or social sciences, which often focus exclusively on descriptive or interpretive or theoretical validity. The validity framework is tailored to the constructivist and utility-oriented nature of DSR. It is consistent with the goals of DSR: to create useful, technically sound, and (whenever appropriate or desirable) generalizable artifacts with clear objectives. The five clearly distinguishable validity types are a well thought-out concept whose origins are rooted in and derived from the DSR logic. Consequently, the validity types can be directly \textbf{mapped to the iterative DSR phases}: Problem Identification (\textit{Instrument Validity}), Design and Evaluation (\textit{Technical Validity}), Design Process (\textit{Design Validity}), Relevance and Theory (\textit{Purpose Validity}), Reflection/Generalization (\textit{Generalization Validity}). This makes the validity framework very practical for the evaluation of artifacts and makes it easier for DSR researchers to structure the evaluation around familiar steps.

\subsection{Mandatory and Flexible Validity Types}
\label{sec:flextyp}

The distinction between the five types of validity is essential, as they each relate to different aspects of the quality and impact of an artifact. Without differentiating between these aspects, critical weaknesses may be overlooked during the evaluation. While it is crucial to understand and differentiate between all five validity types, their application should be flexible and context-dependent. Some projects require all five types, while others may only focus on the most relevant ones. In any case, Evaluation should be systematic, meaning that validity aspects should not be arbitrarily combined or reduced to a single category, such as purpose validity. The structured consideration of all five validity types ensures a rigorous, meaningful and context-sensitive research evaluation.
\\
\\
\textbf{Essential (Mandatory) Validity Types}:
\begin{itemize}
    \item \textit{Instrument Validity} – Ensures that the constructs, measures, and theoretical foundations of research are valid. Without it, the entire study may lack credibility.
    \item \textit{Purpose Validity} – Establishes the relevance and utility of the research. Any study should have a clear purpose and contribute meaningfully to knowledge or practice.
\end{itemize}
\textbf{More Flexible Validity Types} (Depending on Research Context):
\begin{itemize}
    \item \textit{Technical Validity} – Essential for implemented artifacts but less relevant for theoretical work that does not involve direct technological instantiation.
    \item \textit{Design Validity} – Important in design-based research but less critical in exploratory or purely theoretical studies where artifact structuring is not the focus.
    \item \textit{Generalization (External Validity)} – Critical in empirical and applied research but not always required in context-specific or conceptual studies. Some research deliberately focuses on niche scenarios without aiming for generalization.
\end{itemize}

\subsection{Scoring Guidelines for the Five Validity Types in DSR}
\label{sec:scoreguide}

\begin{figure}[htb!]
    \centering
    \includegraphics[width=1.0\textwidth]{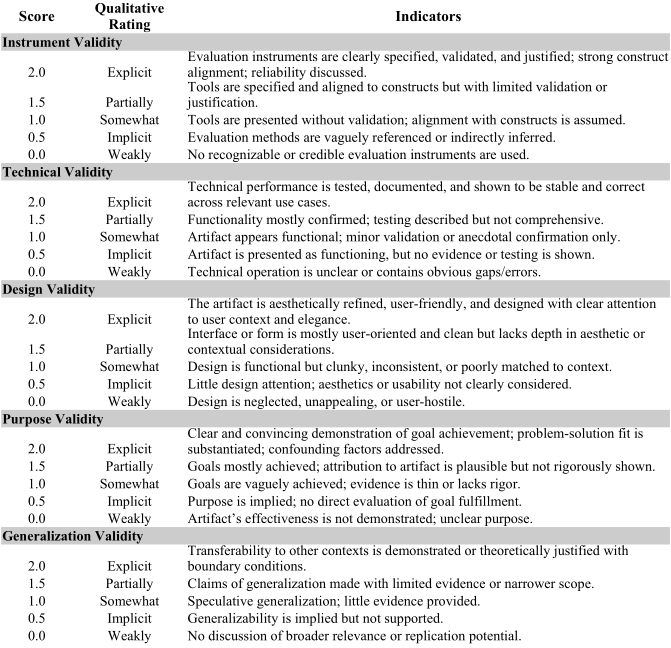}
    \caption{Rubric (scoring grid)}
    \label{fig:rubric}
\end{figure}

Based on the definitions and key questions listed in Sect.~\ref{sec:valtyp}, a rubric (scoring grid) with operationalized indicators for each validity type (i.e., what to look for when evaluating and scoring) was created below. Each validity type is scored in 0.5 steps from 0 (weakly) to 2 (explicit).

These scoring guidelines are designed for the evaluation of artifacts in a wide range of DSR projects. They are generalizable as they are based on the fundamental DSR objectives. The five validity types capture the key evaluation dimensions of artifacts in DSR. They cover the areas of measurement, functionality, usability, problem solving and external relevance. They are not tied to a specific area (e.g. IS, HCI, engineering) and can be applied to software tools, models, methods, frameworks or processes. The guidelines enable a structured assessment across diverse methods. DSR often includes mixed evaluation strategies (experiments, case studies, user tests, simulations, etc.). These guidelines are method-agnostic, focusing instead on what the evaluation demonstrates in terms of validity. For example, instrument validity can be tested in one study with usability logs or in another with surveys - the evaluation logic remains the same. The scoring guidelines are designed to be reproducible and transparent and to support inter-rater reliability. The guidelines make the rationale for scoring explicit and reviewable and help to identify specific strengths or weaknesses in DSR evaluations. The scoring guidelines support both formative and summative evaluation: A DSR team could use this during artifact development (formative feedback) or it could be used by reviewers or researchers for summative assessment. 

\textit{Limitations}: However, some domains might require additional validity dimensions, e.g., ethical validity, sustainability, or stakeholder involvement. For highly technical or mathematical artifacts, proofs or formal models may supplement or replace some aspects (e.g., for technical or purpose validity).

\section{Comparative Evaluation of DSR Frameworks}
\label{sec:discussion}
In this chapter, the scoring guidelines from Sect.~\ref{sec:scoreguide} are used to evaluate the three selected DSR frameworks with regard to the main research question of this paper: To what extent do DSR frameworks support the validity of artifacts? All assessments are based on how explicitly each type of validity is considered and supported in the DSR frameworks, which is finally summarized in an overview in Chap.~\ref{sec:summarized}.

\subsection{Instruments Validity}
\label{sec:instrval}
\textit{Hevner et al. (2004)}, Score (0–2): 0.5 - emphasize rigor in research methods but do not explicitly define instrument validity. While they stress that artifacts should be evaluated using reliable metrics (p. 85), they do not differentiate between validity of the artifact and validity of the instruments used for evaluation. This can lead to methodological weaknesses, as unreliable measurement tools could compromise artifact assessment. As a result, instrument validity remains an implicit concern rather than a structured requirement.

\textit{Peffers et al. (2007)}, Score (0–2): 0 - integrate evaluation as a core phase of their DSR process model, yet they do not explicitly address instrument validity. The framework assumes that research evaluations inherently produce valid findings but does not provide guidance on verifying the accuracy of evaluation instruments. Since measurement errors could affect the validity of conclusions, the absence of explicit considerations for instrument validity leaves a methodological gap, making it one of the least developed aspects.

\textit{Österle et al. (2010) and Benner-Wickner et al. (2020)}, Score (0–2): 0.5 -  emphasize scientific rigor but do not formally distinguish instrument validity as a critical component of evaluation. Although their structured four-phase DSR model (analysis, design, evaluation, diffusion) suggests systematic assessment, it does not explicitly ensure that the instruments used for evaluation are valid. This oversight reinforces the tendency to assume rather than verify measurement validity, making instrument validity among the least developed aspects across all three frameworks.

\subsection{Technical validity}
\label{sec:techval}
\textit{Hevner et al. (2004)}, Score (0–2): 1.5 - acknowledge technical validity through rigorous evaluation of artifact quality. Guideline 3 (Design Evaluation) calls for assessing functionality, reliability, accuracy, and performance (p. 85). However, technical aspects are embedded within broader purpose validity, focusing on effectiveness rather than verifying technical soundness. This poses risks—for instance, learning software may fail not due to purpose misalignment, but due to bugs. Although evaluation methods like experiments and simulations are mentioned (p. 86), technical validity is not defined as a separate requirement. Thus, it is acknowledged but remains only partially addressed.

\textit{Peffers et al. (2007)}, Score (0–2): 1.5 - embed evaluation in their DSR model, requiring researchers to demonstrate artifact effectiveness (p. 56). However, like Hevner et al., they do not distinguish technical validity from purpose or design validity. The emphasis on relevance and utility aligns more with purpose validity. Though feedback loops support refinement, the framework lacks systematic validation methods. This may result in artifacts being judged effective before ensuring technical reliability, posing practical risks. By prioritizing usability and relevance over explicit technical rigor, technical validity is acknowledged but remains only partially considered.

\textit{Österle et al. (2010) and Benner-Wickner et al. (2020)}, Score (0–2): 1.5 - offer a structured DSR process but, like earlier frameworks, do not define technical validity as a separate requirement. They incorporate empirical standards from software engineering~\cite{ralph_empirical_2021}, which strengthens technical validation compared to Hevner et al. and Peffers et al. Still, the framework assumes evaluation inherently covers technical issues without explicitly prioritizing them. Thus, while technical reliability is acknowledged, it remains only partially considered, embedded in broader evaluation rather than treated as a required, standalone step.

\subsection{Design validity}
\label{sec:desgval}
\textit{Hevner et al. (2004)}, Score (0–2): 1 - emphasize artifact utility but do not explicitly define design validity as a distinct evaluation criterion. While they discuss usability, completeness, and functionality as relevant quality attributes (p. 85), they do not provide systematic methods to assess coherence, user experience, or aesthetic quality. The framework prioritizes problem-solving and effectiveness over structured design validation, making design validity underdeveloped and largely assumed rather than explicitly assessed.

\textit{Peffers et al. (2007)}, Score (0–2): 0 - acknowledge artifact usability as part of iterative refinement but do not explicitly define design validity. The DSR process model ensures that artifacts evolve based on feedback loops, but it does not establish specific design evaluation criteria. As a result, usability and aesthetic considerations remain secondary to functionality and problem-solving. Without clear guidance on assessing logical structure, clarity, or user-friendliness, design validity remains one of the least developed aspects.

\textit{Österle et al. (2010) and Benner-Wickner et al. (2020)}, Score (0–2): 1 - emphasize practical relevance and structured evaluation, but design validity is not separately addressed. While their framework improves accessibility and application through a structured DSR process, it does not explicitly differentiate usability or aesthetic considerations from broader artifact evaluation. By focusing on practical implementation rather than formalized design assessment, design validity remains largely underdeveloped, similar to the other frameworks.

\subsection{Purpose validity}
\label{sec:purpval}

\textit{Hevner et al. (2004)}, Score (0–2): 2 - emphasize purpose validity by defining DSR as the creation of artifacts that improve organizational effectiveness and efficiency (p. 76). They state that artifacts must be purposeful, addressing key organizational problems (p. 82), reinforcing that utility and knowledge contribution are inseparable in DSR. The framework ensures feasibility assessment, confirming whether an artifact meets its intended purpose (p. 79).

\textit{Peffers et al. (2007)}, , Score (0–2): 2 - explicitly link DSR to problem-solving and human purpose (p. 55), stating that successful artifacts must meet predefined objectives (p. 46). Their DSR process model prioritizes problem identification and objectives as key research phases, ensuring artifacts are designed with clear intent and evaluated based on their effectiveness (p. 54). This structured approach reinforces the direct connection between problem relevance, utility, and knowledge contribution.

\textit{Österle et al. (2010) and Benner-Wickner et al. (2020)}, Score (0–2): 2 - strongly emphasize practical relevance and societal impact by structuring DSR into analysis, design, evaluation, and diffusion phases, ensuring artifacts serve real-world needs. Benner-Wickner et al. enhance purpose validity further by transforming this structured process into a clear, visually accessible framework, making DSR principles easier to apply and reinforcing the artifact’s practical relevance.

\subsection{Generalization (external validity)}
\label{sec:generalval}

\textit{Hevner et al. (2004)}, Score (0–2): 1 - recognize that DSR contributes to both practice and theory (p. 79) but do not explicitly emphasize external validity. While artifacts are meant to apply to real - world contexts, the framework doesn’t focus on transferability beyond the initial problem space. It assumes utility in one environment implies broader applicability, so generalization is only somewhat addressed—not a key concern.

\textit{Peffers et al. (2007)}, Score (0–2): 1 - emphasize artifact effectiveness in its context but offer no clear guidance on generalizing results. Their process model promotes problem-solving and iterative refinement but lacks mechanisms for testing findings in varied settings. Thus, generalization is secondary to immediate artifact utility — acknowledged, but not emphasized.

\textit{Österle et al. (2010) and Benner-Wickner et al. (2020)}, Score (0–2): 1 - stress practical impact and diffusion as key DSR phases, implying some attention to generalization. However, their real-world focus doesn’t extend to explicitly ensuring external validity. Although diffusion supports broader application, it lacks mechanisms to validate artifact use across domains. Generalization remains a secondary concern.

\section{Summarized assessment of the five validity types in the three DSR frameworks}
\label{sec:summarized}

\begin{figure}[htb!]
    \centering
    \includegraphics[width=1.0\textwidth]{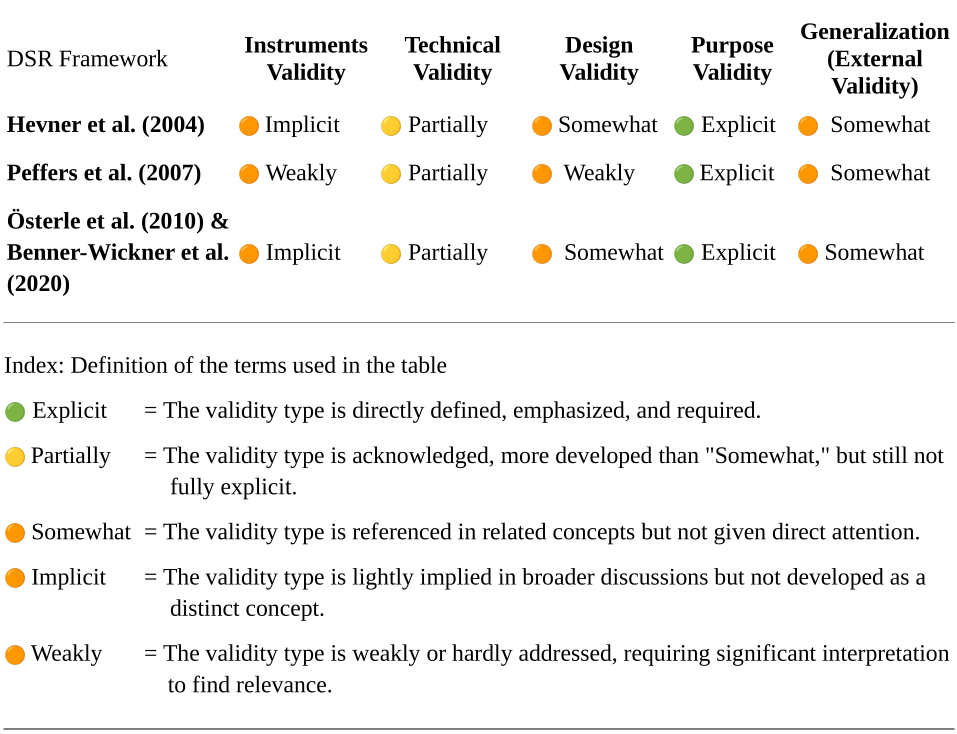}
    \caption{Consideration of the five validity types in three DSR frameworks}
    \label{fig:tabvalidDSR}
\end{figure}

The table in Figure~\ref{fig:tabvalidDSR} provides a summarized overview of how each of the DSR frameworks discussed in chap.~\ref{sec:discussion} addresses the five types of validity. This overview should help researchers and students to quickly identify the strengths, weaknesses and gaps in the validity considerations of each DSR framework. It also leads to the recommendation to seek a more in-depth discussion about the validation of an artifact.

Each validity type is rated using a \textit{qualitative scale}, represented with colored labels, ranging from \textbf{explicit consideration} (\textit{most emphasized}) to \textbf{weak consideration} (\textit{least emphasized}). From this, the following patterns and commonalities can be identified:

\begin{itemize}
    \item Purpose validity is the most developed across all three frameworks. 
    \item Technical validity is acknowledged in all three frameworks, but only partially considered.
    \item Generalization is somewhat addressed, but not strongly emphasized in any framework.
    \item Instrument validity and design validity are the least developed overall.
\end{itemize}

To date, there has been no triangulation of the scoring process (e.g., through independent assessment by multiple reviewers). Therefore, the qualitative assessments from "somewhat" to "weakly" were summarized into a uniform color code (orange). The goal was to highlight the trends in validity support within DSR frameworks. Purpose validity - although marked in green - could (like all other validity types) be more strongly supported within DSR frameworks, particularly with regard to guidance on controlling for and avoiding confounding factors during the evaluation. However, compared to all other validity types, purpose validity is considered the most explicitly addressed of all DSR frameworks.

\section{Conclusion and Future Work}
\label{sec:conclusion}
This paper compares three influential DSR frameworks to assess their support for artifact validity. Five essential validity types were applied to the context of Design Science Research (DSR). The analysis reveals that while purpose validity is explicitly emphasized in all three frameworks, instrument validity and design validity remain the least developed.

Although DSR rigor implicitly encompasses all five validity types, none of the frameworks explicitly define or differentiate them. This poses a risk that mandatory instrument validity may be overlooked, leading to invalid artifacts and undermining research credibility. The biggest weakness arises when students or researchers rely solely on one of the frameworks without recognizing the need for deeper validation studies, particularly regarding instrument validity. Evaluation should not oversimplify validity by subsuming it under a single category, such as purpose validity, but should systematically process each type to ensure comprehensive assessment. 

To address these gaps, this paper provides (a) a comparative overview highlighting strengths and weaknesses in each framework and (b) a revised and extended DSR framework that systematically integrates five validity types with definitions and examples and scoring guidelines. This structured approach enhances the rigor and reliability of DSR research.

Additionally, the lack of explicit emphasis on design validity, distinct from technical and other validity types, is notable given that design is central to DSR. While design validity may not be as critical as instrument validity, greater focus on user-centric criteria - such as simplicity, taste, style, and elegance - could enhance artifact adoption and usability. The revised framework proposed in this paper acknowledges the role of such subjective and aesthetic factors and can be adapted to different types of artifacts (e.g., software, processes, physical tools), ensuring broad applicability.

\textbf{Further research} could validate the revised and extended DSR framework through case studies and establish formal methods for instrument validity, and also include ethical and ecological validity as standalone measurement instruments. Exploring the impact of weak instrument validity and AI-driven assessments could enhance research credibility. Comparative analyses across domains may reveal best practices for ensuring validity. Addressing these issues will refine artifact evaluation, strengthening DSR’s rigor and impact.

% ---- Bibliography ----
\bibliographystyle{plain}  % we recommend the plain bibliography style
\bibliography{literature.bib}       % xampl.bib comes with BibTeX
\end{document}